\documentclass[twocolumn, graphics, floatfix, a4paper, aps, prb,
superscriptaddress, longbibliography, showpacs,citeautoscript]{revtex4-1}

\usepackage[english]{babel}

\usepackage[letterpaper,top=2cm,bottom=2cm,left=1.5cm,right=1.0cm,marginparwidth=1.75cm]{geometry}

\usepackage{amsmath, bm}
\usepackage{graphicx}
\usepackage{xcolor}
\usepackage{bbold}
\usepackage[colorlinks=true, allcolors=blue]{hyperref}

\newcommand{\bra}[1]{\langle#1|}
\newcommand{\ket}[1]{|#1\rangle}
\newcommand{\braket}[2]{\langle#1|#2\rangle}

\renewcommand{\vec}[1]{\boldsymbol{#1}}

\newcommand{\ave}[1]{\left\langle #1 \right\rangle}

\newcommand{\crea}[1]{{#1}^{\dag}}
\newcommand{\anni}[1]{{#1}^{\vphantom{\dag}}}

\begin{document}

\title{Interplay of local and global quantum geometry in the stability of flat-band superfluids} 
\author{Kukka-Emilia Huhtinen}
\email{khuhtinen@phys.ethz.ch}
\affiliation{
Institute for Theoretical Physics, ETH Zurich, 8093 Zürich, Switzerland
}
\author{Matteo Dürrnagel}
\affiliation{
Institute for Theoretical Physics, ETH Zurich, 8093 Zürich, Switzerland
}
\affiliation{Institut für Theoretische Physik und Astrophysik, Universität Würzburg, 97074
Würzburg, Germany}
\author{Valerio Peri}
\altaffiliation{Present address: Hexagon Innovation Hub GmbH, Heinrich-Wild-Strasse, 9435 Heerbrugg, Switzerland}
\affiliation{Department of Physics, IQIM, California Institute of Technology, Pasadena, CA 91125, USA.}
\author{Sebastian D. Huber}
\affiliation{
Institute for Theoretical Physics, ETH Zurich, 8093 Zürich, Switzerland
}
\date{\today}

\begin{abstract}
Quantum geometry strongly impacts physical properties in flat-band
systems. We consider its role in bosonic condensation and superfluidity on flat bands,
and show that the superfluid weight has an important contribution
proportional to the condensate quantum metric. Based on this result, we uncover conditions under which flat-band superfluidity is unlikely. For instance, we find
that stable flat-band superfluidity in a two-dimensional system
requires at least three bands within Bogoliubov theory. Because the quantum geometry at the condensation momentum plays a
disproportionately large role, a large integrated quantum metric is
not sufficient for flat-band superfluidity, but how the quantum
metric is distributed in the Brillouin zone is crucial.
\end{abstract}

\maketitle

\section{Introduction}

The importance of quantum geometry in the physics of multiband systems
is well-established. The quantum geometric tensor is often used to describe the geometry of Bloch states. Its imaginary
part, the Berry curvature, has been extensively
studied~\cite{Xiao2010}, for instance in the context of the anomalous
Hall effect~\cite{Nagaosa2010}. The real part, dubbed the quantum
metric~\cite{Provost1980}, has attracted increased attention in recent
years, and is involved in a variety of phenomena~\cite{Torma2022,Jiang2025a,Yu2025,Liu2025}. For instance, the
quantum metric influences the superfluid weight and
effective masses in fermionic flat-band
systems~\cite{Peotta2015,Julku2016,Tovmasyan2016,Liang2017,Iskin2019,Iskin2019b,Verma2021,Kitamura2022,Huhtinen2022,Herzog-Arbeitman2022,Tam2024,Iskin2024,Jiang2024}, and nontrivial quantum geometry 
of the Bloch states can enable superconductivity even when single particles are localized.

An interesting question is whether condensation of bosons in a flat
band is made possible by similar geometric effects. Recent research
has uncovered relationships between 
physical properties of Bose-Einstein condensates and quantum geometric
quantities~\cite{Julku2021a,Julku2021b,Iskin2023,Lukin2023,Julku2023,Jalali-mola2023}. In particular, the quantum metric influences the
speed of sound~\cite{Julku2021a,Julku2021b,Julku2023} as well as the effective mass of superfluid
carriers~\cite{Iskin2023} in flat bands. Nontrivial quantum geometry can therefore enable
a somewhat counter-intuitive phenomenon: bosonic condensation and superfluidity in flat
bands, where all states are degenerate at the single-particle
level. Stable condensation is indeed predicted in some simple
flat-band models such as the kagome lattice~\cite{Huber2010,You2012,Julku2021a}.

A central quantity when it comes to superfluidity is the
superfluid weight. The expressions for the superfluid weight have been
derived within Bogoliubov
theory~\cite{Julku2021a,Julku2021b,Julku2023} as well as using a Popov
hydrodynamic formalism~\cite{Lukin2023}. In this work, we find that
the superfluid weight at low interactions has an important
contribution proportional to a so-called condensate quantum metric,
which reduces to the usual quantum metric at the condensation momentum
under certain conditions. In some systems such as the kagome lattice,
this relationship holds even outside the low-interaction regime within
Bogoliubov theory. In addition, fluctuations lead to a further contribution to
the superfluid weight, which depends on the geometry of Bloch states
throughout the Brillouin zone.

Based on the disproportionate role of the quantum metric at a single
momentum compared to other momenta, we determine conditions under
which flat-band superfluidity is unlikely to occur. For instance, in
two-dimensional systems, we predict that superfluidity is difficult to
achieve in two-band flat-band models, as well as at high-symmetry
points which are at time-reversal invariant momenta. In strong
contrast to fermionic superconductors, where any geometrically
nontrivial band is predicted to be compatible with superconductivity,
Bose-Einstein condensation and superfluidity is not necessarily possible even in highly
nontrivial bands.

\section{Bogoliubov theory}

We consider the Bose-Hubbard model on a lattice,
    \begin{equation}
    \begin{aligned}
    H =& -\sum_{i\alpha,j\beta} (t_{i\alpha,j\beta} -\mu\delta_{i\alpha,j\beta} ) 
    \crea{b}_{i\alpha} \anni{b}_{j\beta}\\
    +& \frac{U}{2}\sum_{i\alpha} \crea{b}_{i\alpha} \crea{b}_{i\alpha} \anni{b}_{i\alpha} \anni{b}_{i\alpha},
    \end{aligned}
    \end{equation}
where $\crea{b}_{i\alpha}$ creates a boson at site $i\alpha$, with $i$
labeling the unit cell and $\alpha$ one of the $M$ orbitals within the
unit cell. Particles can hop from site $j\beta$ to
site $i\alpha$ with amplitude $t_{i\alpha,j\beta}$ and experience an
on-site repulsive interaction $U$. The chemical potential is $\mu$. 

In Bogoliubov theory, we expand $b_{i\alpha} =
\sqrt{n_0}\varphi_0(\vec{r}_{i\alpha}) + c_{i\alpha}$, where $n_0$ is the
density of condensed particles per unit cell and condensation is
assumed to occur in the state $\varphi_0(\vec{r}_{i\alpha})$. The
operators $\crea{c_{i\alpha}}$ create fluctuations and follow bosonic
commutation relations. In this work, we assume that $\varphi_0(\vec{r}_{i\alpha}) =
\braket{\alpha}{\varphi_0}e^{i\vec{k_c}\cdot \vec{r}_{i\alpha}}$ is a
flat-band eigenstate corresponding to a well-defined condensation
momentum $\vec{k_c}$. We note that on a flat band, a multitude of
states can minimize $E_{\rm MF}$, many of which do not fulfill this
assumption, and numerical computations of the zero-point energy are
typically necessary to determine which is the most
favorable~\cite{You2012,realspace}. For many simple models, such as
the kagome lattice~\cite{You2012} and the Tasaki model considered in
Ref.~\onlinecite{realspace}, the most favorable state is a Bloch
function or a superposition of a few Bloch states, and it is possible
to enlarge the unit cell so that $\varphi_0$ corresponds to a
well-defined $\vec{k_c}$.  

We ignore terms that are higher than quadratic in the fluctuation
operators, and obtain the following Fourier transformed Hamiltonian:
\begin{equation}
  \begin{aligned}
  \frac{H}{N} &= \frac{1}{2N}\sum_{\vec{k}} \vec{\crea{\psi}}_{\vec{k}}
  H_B(\vec{k}) \vec{\anni{\psi}}_{\vec{k}} + E_{\rm MF} + C, \\
  H_B(\vec{k}) &= \begin{pmatrix}
    H_{\vec{k_c}+\vec{k}} - \mu_{\rm eff} & \Delta \\
    \Delta^{\dag} & H_{\vec{k_c}-\vec{k}}^*-\mu_{\rm eff}
  \end{pmatrix},\\
  E_{\rm MF} &= n_0\bra{\varphi_0}H_{\vec{k_c}}\ket{\varphi_0} -\mu
  n_0 + \frac{Un_0^2}{2}\sum_{\alpha}|\braket{\alpha}{\varphi_0}|^4,
  \\
  C &= -\frac{1}{2N}{\rm Tr}\left[\sum_{\vec{k}}H_{\vec{k}} \right] +
  \frac{\mu}{2} - Un_0.
  \end{aligned}
\end{equation}
Here, $[H_{\vec{k}}]_{\alpha\beta} = \sum_j
t_{0\alpha,j\beta}e^{-i\vec{k}\cdot(\vec{r}_{0\alpha}-\vec{r}_{j\beta})}$
is the Fourier transformation of the single-particle Hamiltonian, and
$\crea{\vec{\psi}}_{\vec{k}} =
(\crea{c}_{\vec{k_c}+\vec{k},1},\ldots,\crea{c}_{\vec{k_c}+\vec{k},M},\anni{c}_{\vec{k_c}-\vec{k},1},\ldots,\anni{c}_{\vec{k_c}-\vec{k},M})$,
with $c_{\vec{k}\alpha} =
(1/\sqrt{N})\sum_{\vec{k}}c_{i\alpha}e^{-i\vec{k}\cdot\vec{r}_{i\alpha}}$. We
define $[\Delta]_{\alpha\beta} =
Un_0\braket{\alpha}{\varphi_0}^2\delta_{\alpha\beta}$ and $[\mu_{\rm
    eff}]_{\alpha\beta} =
(\mu-2Un_0|\braket{\alpha}{\varphi_0}|^2)\delta_{\alpha\beta}$. 

The state $\ket{\varphi_0}$ is determined by minimizing the mean-field
energy $E_{\rm MF}$ under the constraint
$\sum_{\alpha}|\braket{\alpha}{\varphi_0}|^2=1$. As mentioned above, we will assume
throughout that $\ket{\varphi_0}$ is a Bloch state corresponding to
$\vec{k_c}$ on the lowest band. The condensation momentum $\vec{k_c}$
is chosen to be the one that gives the lowest possible $E_{\rm MF}$. Since
we focus on flat-band condensation,
$\bra{\varphi_0}H_{\vec{k_c}}\ket{\varphi_0}=\epsilon_0$, where
$\epsilon_0$ is the momentum-independent energy of the flat
band. The mean-field energy is therefore minimized when
$\sum_{\alpha}|\braket{\alpha}{\varphi_0}|^4$ is minimized. If such a
state is possible, this occurs when
$|\braket{\alpha}{\varphi_0}|^2=1/M$. Such a uniform-density state
exists in many simple flat-band models, and we will assume for
analytical calculations that $|\braket{\alpha}{\varphi_0}|^2=1/M$.

The computation of thermodynamic quantities is typically most
conveniently achieved by determining Bogoliubov boson operators that
diagonalize $H_B$. Because these operators must follow bosonic commutation
relations, we need to determine $E_n^s(\vec{k})$ and
$\ket{n^s(\vec{k})}$, $s=\pm$, $n\in[1,M]$, such that
$H_B(\vec{k})\ket{n^s(\vec{k})} = E_n^s(\vec{k}) \ket{n^s(\vec{k})}$ and
$\bra{n^s(\vec{k})}\gamma_z\ket{m^{s'}(\vec{k})} =
s\delta_{nm}\delta_{ss'}$, with $\gamma_z=\sigma_z\otimes
\mathbb{1}_{M}$ and $\mathbb{1}_M$ the $M\times M$ identity matrix. In
other words, we carry out paraunitary diagonalization of
$H_B$~\cite{Colpa1978,Colpa1986a,Colpa1986b} to determine $E_n^s$ and
$\ket{n^s}$. In practice, this is achieved by diagonalizing the
non-Hermitian matrix $\gamma_zH_B(\vec{k})$, which in our case has
real eigenvalues and is diagonalizable at all momenta except
$\vec{k}=0$, due to the existence of an improper zero mode related to
the condensation~\cite{Colpa1986b,Xu2020}.

For a given $\ket{\varphi_0}$ and total density $n_{\rm tot}$, the
condensate fraction $n_0$ needs to be solved self-consistently. To do
so, we compute the excitation fraction $n_{\rm ex}$ for an initial
guess of $n_0$, which in the $T\to 0^+$ limit reads $n_{\rm ex} =
(1/2N)\sum_{\vec{k}\neq 0,n}(\braket{n^-(\vec{k})}{n^-(\vec{k})}
-1)$. We then obtain a new value for $n_0$ from $n_{\rm tot} =
n_0+n_{\rm ex}$. This procedure is repeated until $n_0$
converges. Note that, as we focus on flat-band condensation,
$\ket{\varphi_0}$ is independent of $n_0$. However, in a dispersive band, the condensation momentum could shift with $n_0$, and it
would also be necessary to update $\ket{\varphi_0}$ during the
computation.

The central quantity we consider is the superfluid weight. In linear
response theory, it can be defined as $D_{\mu\nu} =
\lim_{\mathbf{q}\to 0}\lim_{\omega\to 0}
K_{\mu\nu}(\omega,\mathbf{q})$. Here $K_{\mu\nu}$ is the
current-current response function, $j_{\mu}(\omega,\mathbf{q}) =
K_{\mu\nu}(\omega,\mathbf{q}) \vec{A}_{\nu}(\omega,\mathbf{q})$, where
$j_{\mu}(\omega,\mathbf{q})$ is the current density resulting from a
vector potential $\vec{A}(\omega,\mathbf{q})$. In the usual linear
response theory,
\begin{equation}
  \begin{aligned}
    K_{\mu\nu} &= \ave{T_{\mu\nu}} - i\int_0^{\infty} dt\: e^{i\omega
      t}\ave{[j_{\mu}^p(\mathbf{q},t),j_{\nu}^{p}(-\mathbf{q},0) ]},
    \\
    T_{\mu\nu} &= \sum_{\vec{k}}\crea{b_{\vec{k}}}
    \partial_{\mu}\partial_{\nu} H_{\vec{k}} \anni{b_{\vec{k}}}, \\
    j_{\mu}^p(\mathbf{q}) &=
    \sum_{\vec{k}}\crea{b_{\vec{k}+\mathbf{q}}}
    \partial_{\mu}H_{\vec{k}+\mathbf{q}/2}\anni{b_{\vec{k}}}, \label{eq.linresp} 
  \end{aligned}
\end{equation}
where we use the shorthand
$\partial_{\mu}\equiv\partial/\partial_{k_{\mu}}$. 
After performing the substitution $b_{\vec{k}\alpha} =
\sqrt{n_0}\braket{\alpha}{\varphi_0}\delta_{\vec{k},0}+c_{\vec{k}\alpha}$,
we can divide the total superfluid weight $D$ into three
contributions: $D_1$, which involves only the condensate but none of
the operators $c_{\vec{k}\alpha}$, $D_2$ which involves both the condensate and
fluctuations, and $D_3$ which involves only the fluctuation operators
$c_{\vec{k}\alpha}$. 

Because $n_0$, and potentially $\ket{\varphi_0}$, need to be solved
self-consistently, Eq.~\eqref{eq.linresp} does not actually give the
full superfluid weight within Bogoliubov theory. Instead, additional
terms need to be added to account for the dependence of
$n_0$~\cite{Julku2023} and $\ket{\varphi_0}$ on the vector potential $\vec{A}$. These
terms are important to ensure the electromagnetic gauge-invariance of
the superfluid weight. In addition, within Bogoliubov theory, $n_0$
does not need to be exactly at a minimum of the thermodynamic
potential $\Omega$~\cite{Yukalov2006a,Yukalov2006b,Andersen2004,Griffin1996}, which leads to further contributions to the
superfluid weight~\cite{Julku2023}. These latter contributions should be small
when Bogoliubov theory is applicable, since a $\partial\Omega/\partial n_0$
significantly deviating from zero would indicate that the Bogoliubov theory is
likely a bad approximation. In this work, we group all these
additional contributions and label them $D_{\rm cor}$.
The total superfluid weight can then be written as
$D=D_1+D_2+D_3+D_{\rm cor}$.

As an example model, we consider the kagome
lattice, pictured in Fig.~\ref{fig.kagome}a. We choose the sign of
hopping amplitudes so that the lowest band is flat, with a
quadratic band touching at the $\Gamma$ point (see
Fig.~\ref{fig.kagome}b). Both Bloch functions at the $\Gamma$ and $K$
points of the Brillouin zone fulfill the uniformity assumption
$|\braket{\varphi_0}{\alpha}|^2=1/M$, and thus minimize the mean-field
energy. However, once fluctuations are included, the $K$ point is more
favorable for condensation~\cite{You2012}, and we thus focus on condensation
at this momentum. The numerically computed $D_1$, $D_2$, $D_3$ and
$D_{\rm cor}$ are pictured in Fig.~\ref{fig.kagome}c-d. In the following,
we show that $D_1$ and $D_2$ in particular relate to the geometry of
Bloch states at $\vec{k_c}$, and consider how this impacts the
possibility of superfluidity in a flat band. 

\section{Superfluid weight in a flat band}

\subsection{Condensate quantum metric}

\begin{figure}
\centering
\includegraphics[width=\columnwidth]{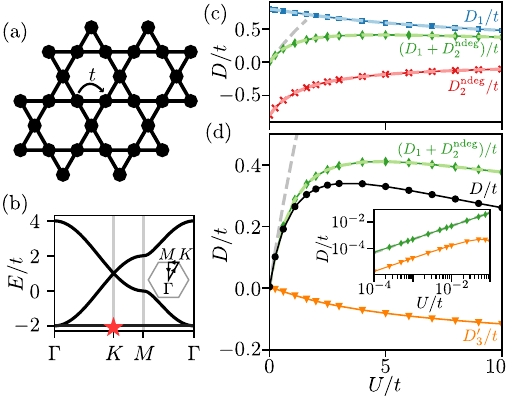}
\caption{(a) Kagome lattice and (b) corresponding band structure. In
  the kagome lattice, condensation occurs at the $K$ point (indicated
  by the red star), where the two dispersive bands above the flat band
  are degenerate. (c) Contributions $D_1$ (blue squares), $D_2^{\rm ndeg}$
  (red crosses) and their sum $D_1+D_2^{\rm ndeg}$ (green diamonds) as a function of
  the on-site interaction $U$. In the kagome lattice, the superfluid
  weight is diagonal and proportional to the identity matrix, so we
  plot only the $xx$ component. The predictions relating these
  contributions to the condensate quantum metric, obtained from
  Eqs.~\eqref{eq.d1} and~\eqref{eq.arbitrary} (dashed blue, red and green lines) agree perfectly with our numerical results at all
  interactions. The more general relationship Eq.~\eqref{eq.d2}, valid
  also when the dispersive bands are not degenerate at $\vec{k_c}$,
  captures only the low-interaction behavior (gray dashed line). The
  dashed lines are plotted using the numerically obtained values for
  $n_0$ at each interaction. (d) Contributions $D_1+D_2^{\rm ndeg}$
  (green diamonds), $D_3'$ (orange triangles), and total superfluid weight $D=D_1+D_2^{\rm ndeg}+D_3'$ (black dots) as a function of the interaction strength. Inset: $D_1+D_2^{\rm ndeg}$ (green diamonds) and $D_3'$ (orange triangles) at low $U$. The gray dashed line is obtained from the low-$U$ prediction Eq.~\eqref{eq.d2} for $D_1+D_2^{\rm ndeg}$, and agrees well with the numerical data. In the kagome lattice, $D_3'$ is positive at very small $U$, and changes sign at approximately $U\approx 0.13t$. }
\label{fig.kagome}
\end{figure}

We first focus on the contribution $D_1$, arising from the condensate
alone, and $D_2$, involving both the condensate and fluctuations. In
the following, we consider condensation in a single flat band, labeled
$\overline{n}$, at momentum $\vec{k_c}$. We assume that  
there are no band touchings with dispersive bands exactly at
$\vec{k_c}$, although band touchings can exist elsewhere. Under these
conditions, $D_1$ takes the simple form~\cite{Julku2021b}
\begin{equation}
  \begin{aligned}
    D_{1,\mu\nu} &= n_0\bra{\varphi_0}\partial_{\mu}\partial_{\nu} H_{\vec{k_c}}\ket{\varphi_0} \\
    &=n_0\sum_{m> \overline{n}} [
      (\epsilon^m_{\boldsymbol{k_c}}-\epsilon_0)
      \langle \partial_{\mu}\varphi_0 | m_{\vec{k_c}} \rangle\langle
      m_{\vec{k_c}} | \partial_{\nu} \varphi_0 \rangle +
      \mu\leftrightarrow\nu]. \label{eq.d1}
  \end{aligned}
\end{equation}
Since we assumed there are no band touchings at $\vec{k_c}$, all terms
involving derivatives of the band dispersion vanish. The derivative of
$\ket{\varphi_0}$ is defined as the derivative of the Bloch state
$\ket{\overline{n}_{\vec{k}}}$ of the lowest band at $\vec{k_c}$,
$\ket{\partial_{\mu}\varphi_0}\equiv
\ket{\partial_{\mu}\overline{n}_{\vec{k}}}\big|_{\vec{k}=\vec{k_c}}$. 

The contribution $D_2$ is given by~\cite{Julku2021b}
\begin{equation}
  \begin{aligned}
  &D_{2,\mu\nu} = -n_0\lim_{\vec{k}\to 0} \sum_{ms}\\
  &\frac{\bra{\vec{\varphi_0}}
    \gamma_z\partial_{\mu}H_B(-\vec{k}/2)\ket{m^s(-\vec{k})}
    \bra{m^s(-\vec{k})}
    \gamma_z\partial_{\nu}H_B(-\vec{k}/2)\ket{\vec{\varphi_0}}}{E_m^+(-s\vec{k})},
  \end{aligned}
\end{equation}
where $\ket{\vec{\varphi_0}} = (\ket{\varphi_0},\ket{\varphi_0^*})^T$.

We obtain $E_n^s(\vec{k})$ and $\ket{n^s(\vec{k})}$ for $\vec{k}\neq
0$ at low $\Delta=Un_0/M$ from perturbation theory (see
Appendix~\ref{app.perturbation}). For $m>\overline{n}$, the states are
labeled so that $\lim_{U\to 0}\ket{m^+(\vec{k})} = \ket{+}\otimes
\ket{m_{\vec{k_c}+\vec{k}}}$ and $\lim_{U\to 0} \ket{m^-(\vec{k})} =
\ket{-}\otimes\ket{m^*_{\vec{k_c}-\vec{k}}}$, where $\ket{\pm}$ are
the eigenvectors of $\sigma_z$ with eigenvalues $\pm
1$. In the same limit, the eigenvectors $\ket{\overline{n}^s_{\vec{k}}}$, are
superpositions of $\ket{+}\otimes\ket{\overline{n}_{\vec{k_c}+\vec{k}}}$ and
$\ket{-}\otimes\ket{\overline{n}^*_{\vec{k_c}-\vec{k}}}$, 
since both of these states correspond to the same eigenvalue $0$ of
$H_B(\vec{k})$ at $U=0$.

The contribution $D_2$ can be split into two parts, $D_{2,\mu\nu} =
D_{2,\mu\nu}^{\rm ndeg} + D_{2,\mu\nu}^{\rm deg}$, with
    \begin{equation}
    \begin{aligned}
    &D_{2,\mu\nu}^{\rm ndeg} =
    -n_0\lim_{\vec{k}\to 0} \sum_{m>\overline{n}}\sum_s \\
    &\frac{\bra{\vec{\varphi_0}} \gamma_z\partial_{\mu}H_{B}(-\vec{k}/2) \ket{m^s(-\vec{k})}  \bra{m^s(-\vec{k})} \gamma_z\partial_{\nu}H_B(-\vec{k}/2) \ket{\vec{\varphi_0}} } {E_m^+(-s\vec{k})}.
    \end{aligned}
    \end{equation}
    The part $D_{2}^{\rm ndeg}$ contains contributions involving only
    $\ket{m^s}$ with $m>\overline{n}$, i.e. eigenvectors that do not
    correspond to the lowest mode.

For a condensate in a flat band fulfilling the uniformity condition
$|\braket{\alpha}{\varphi_0}|^2=1/M$, it can be shown that at low
$Un_0^2/M$ (see appendix~\ref{app.d2})
\begin{equation}
    \begin{aligned}
  D_{2,\mu\nu}^{\rm ndeg} 
  &= -D_{1,\mu\nu} + \frac{2Un_0^2}{M} \mathcal{M}^D_{\mu\nu}(\vec{k_c}), \\
  \mathcal{M}^D_{\mu\nu}(\vec{k_c}) &=
  2M\sum_{\alpha} {\rm
    Re}[\braket{\partial_{\mu}\varphi_0}{\alpha}\braket{\alpha}{\varphi_0}]
  {\rm
    Re}[\braket{\varphi_0}{\alpha}\braket{\alpha}{\partial_{\nu}\varphi_0}]. \label{eq.d2} 
\end{aligned}
\end{equation}
This relationship holds when
$Un_0^2\ll|\epsilon_{\vec{k_c}}^m-\epsilon_0|$ for all
$m>\overline{n}$, i.e. the flat band is isolated from other bands at
$\vec{k_c}$. Since $D_2^{\rm ndeg}$ only involves the momentum $\vec{k_c}$, band
touchings can still occur elsewhere without affecting the validity of
this equation. In the kagome lattice, for instance,
condensation occurs at the $K$-point, where there are no band
touchings, and Eq.~\eqref{eq.d2} indeed captures the slope of
$D_1+D_2^{\rm ndeg}$ at low $U$ (see Fig.~\ref{fig.kagome}c-d). 

We refer to the central quantity $\mathcal{M}^D(\vec{k_c})$ as the condensate
quantum metric. It is not the usual quantum metric, but a
basis-invariant quantity, i.e. independent of $\vec{\delta}_{\alpha}$,
defined as the position of orbital $\alpha$ with reference to the unit
cell center $\vec{r}_i$, $\vec{r}_{i\alpha}=\vec{r}_i +
\vec{\delta}_{\alpha}$. It reduces to the quantum metric under
specific conditions. Noticing that ${\rm
  Re}[\braket{\partial_{\mu}\varphi_0}{\alpha}\braket{\alpha}{\varphi_0}]
= \sum_{m}{\rm
  Re}[\braket{\partial_{\mu}\varphi_0}{m_{\vec{k_c}}}\braket{m_{\vec{k_c}}}{\alpha}\braket{\alpha}{\varphi_0}]
= \frac{1}{M}{\rm Re}[\braket{\partial_{\mu}\varphi_0}{\varphi_0}] +
\sum_{m>\overline{n}} {\rm
  Re}[[\braket{\partial_{\mu}\varphi_0}{m_{\vec{k_c}}}\braket{m_{\vec{k_c}}}{\alpha}\braket{\alpha}{\varphi_0}]]$
and that
$\braket{\partial_{\mu}\varphi_0}{\varphi_0}=-\braket{\varphi_0}{\partial_{\mu}\varphi_0}$
is purely imaginary, we can rewrite
\begin{equation}
  \begin{aligned}
    \mathcal{M}^D_{\mu\nu}(\vec{k_c})=2M\sum_{\alpha}\sum_{m,l>\overline{n}}
            &{\rm
               Re}[\braket{\partial_{\mu}\varphi_0}{m_{\vec{k_c}}}\braket{m_{\vec{k_c}}}{\alpha}\braket{\alpha}{\varphi_0}]\times
    \\ \times&{\rm
      Re}[\braket{\varphi_0}{\alpha}\braket{\alpha}{l_{\vec{k_c}}}\braket{l_{\vec{k_c}}}{\partial_{\nu}\varphi_0}]. \label{eq.md_reformulated}
  \end{aligned}
\end{equation}
When all of $\ket{\varphi_0}$, $\ket{m_{\vec{k_c}}}$ and
$\ket{\partial_{\mu}\varphi_0}$ are real,
$\mathcal{M}^{D}_{\mu\nu}(\vec{k_c})=\sum_{m>\overline{n}}
\braket{\partial_{\mu}\varphi_0}{m_{\vec{k_c}}}\braket{m_{\vec{k_c}}}{\partial_{\nu}\varphi_0}+\mu\leftrightarrow\nu\equiv\mathcal{M}^0_{\mu\nu}(\vec{k_c})$
is the usual quantum metric. More generally,
\begin{equation}
  \begin{aligned}
    \mathcal{M}^D_{\mu\mu}(\vec{k_c}) &\leq
    2M\sum_{\alpha}\sum_{m,l>\overline{n}} \braket{\partial_{\mu}\varphi_0}{m_{\vec{k_c}}}
      \braket{m_{\vec{k_c}}}{\alpha}\braket{\alpha}{\varphi_0}\times\\
      &\hphantom{\leq
    2M\sum_{\alpha}\sum_{m,l\notin\mathcal{B}}}\times
      \braket{\varphi_0}{\alpha}\braket{\alpha}{l_{\vec{k_c}}}\braket{l_{\vec{k_c}}}{\partial_{\mu}\varphi_0}
    \\
    =& 2\sum_{m>\overline{n}} \braket{\partial_{\mu}\varphi_0}{m_{\vec{k_c}}}
      \braket{m_{\vec{k_c}}}{\partial_{\mu}\varphi_0}
      =\mathcal{M}^0_{\mu\mu}(\vec{k_c}), \label{eq.inequality}
  \end{aligned}
\end{equation}
where we have used that for a complex number $x$, ${\rm Re}[x]{\rm
  Re}[x^*] = {\rm Re}[x]^2 \leq |x|^2$. The diagonal components of the
condensate quantum metric are thus always bounded from above by the
diagonal components of the usual quantum metric.

The inequality of Eq.~\eqref{eq.inequality} evokes the concept of minimal
quantum metric relevant in fermionic flat-band
superconductors~\cite{Huhtinen2022}. In a fermionic system, the superfluid weight
on an isolated flat band with uniform pairing is proportional to the
minimal quantum metric, which is the integrated quantum metric
computed in the basis where its diagonal components are as small as
possible. The minimal quantum metric is, by definition, 
basis-invariant. In our case, the superfluid weight has an important
contribution proportional to the condensate quantum metric, which is
a basis-invariant quantum geometric property of the condensate state
$\ket{\varphi_0}$. When there exists a basis where $H_{\vec{k_c}}$,
$\ket{\varphi_0}$, and $\ket{\partial_{\mu}\varphi_0}$ can all be made
simultaneously real, $\mathcal{M}^D(\vec{k_c})$ will reduce to $\mathcal{M}^0(\vec{k_c})$ {\it in this particular basis}. In any other basis, Eq.~\eqref{eq.inequality} still holds, and the diagonal components of $\mathcal{M}^0(\vec{k_c})$ are larger or equal to those of $\mathcal{M}^D(\vec{k_c})$, which matches $\mathcal{M}^0(\vec{k_c})$ in the real basis. This means that
$\mathcal{M}^D(\vec{k_c})$ is equal to the quantum metric computed in a
basis that makes the diagonal components as small as possible. In
contrast to fermionic systems, where the quantum metric is integrated
before minimization, here we consider only the metric at
$\vec{k_c}$. In cases where $H_{\vec{k_c}}$, $\ket{\varphi_0}$ and
$\ket{\partial_{\mu}\varphi_0}$ cannot be made simultaneously real, the
inequality~\eqref{eq.inequality} might not saturate in any basis, but
the quantum metric itself still gives an upper bound for
$\mathcal{M}^D_{\mu\mu}(\vec{k_c})$. 

In models with $C_2\mathcal{T}$ symmetry, there exists a basis where
the Hamiltonian is real throughout the Brillouin zone~: it is then
possible to construct real Bloch functions with real derivatives. Then
the reality conditions are fulfilled and $[D_1+D_2^{\rm
    ndeg}]_{\mu\nu} = (2Un_0^2/M) \mathcal{M}^0(\vec{k_c})$ in the real
basis. In fact, in such a basis, $D_{2,\mu\nu}^{\rm ndeg}$ can be
written at arbitrary interactions in the form (see
Appendix~\ref{app.d2_arbitrary}): 
\begin{equation}
\begin{aligned}
    D_{2,\mu\nu}^{\rm ndeg} &= -n_0\sum_{m>\overline{n}}
      \frac{(\epsilon_{\vec{k_c}}^m -
        \epsilon_0)^2}{\epsilon_{\vec{k_c}}^m - \epsilon_0 + 2\Delta}
      \times\\
      &\times \left[
      \braket{\partial_{\mu}\varphi_0}{m_{\vec{k_c}}}\braket{m_{\vec{k_c}}}{\partial_{\nu}\varphi_0}
      + \mu\leftrightarrow\nu\right]. \label{eq.arbitrary}
\end{aligned}
\end{equation}
In the particular case where $\epsilon_{\vec{k_c}}^m-\epsilon_0 =
E_{\rm gap}$ for all bands $m\notin\mathcal{B}$, $D_{2,\mu\nu}^{\rm
  ndeg} = -n_0E_{\rm gap}^2/(E_{\rm gap} + 2\Delta)
\mathcal{M}_{\mu\nu}^0(\vec{k_c})$ at all interactions. The kagome lattice is an
example of such a model, since condensation occurs at the $K$-point,
where the two dispersive bands touch. As shown in
Fig.~\ref{fig.kagome}c, Eq.~\eqref{eq.arbitrary} is in perfect
numerical agreement with our results when using the self-consistently
solved $n_0$. Note that it is necessary to take the $U$-dependence of $n_0$ into account to obtain the correct $U$-dependence
of $D_{2,\mu\nu}^{\rm   ndeg}$. The approximation $\Delta \approx
n_0(U=0)U/M$ would only give an approximation of the low-$U$
behavior. 

We note that it was already known from previous literature that $D_1+D_2$ vanishes in the limit of zero interaction~\cite{Julku2021b,Lukin2023}; this does not mean $D_1+D_2$ vanishes at nonzero $U$. Instead, as shown by Eq.~\eqref{eq.d2}, $D_1+D_2$ has a contribution linear in $U$ proportional to the condensate quantum metric. As explained below, this contribution is crucial to stabilize flat-band superfluidity.

\subsection{Other contributions}

As shown in the previous section, when condensation occurs in a single
flat band, $D_1+D_{2,\mu\nu}^{\rm ndeg}$ constitutes a basis-invariant
contribution proportional to a condensate quantum metric
$\mathcal{M}^D(\vec{k_c})$. This quantity is positive semidefinite (see
Appendix.~\ref{app.positive}), and $D_1+D_{2,\mu\nu}^{\rm ndeg}$ is
thus positive when condensing into a uniform-density Bloch state on a
single band. Since the whole superfluid weight is basis-independent,
the remaining contributions $D_3'=D_{2}^{\rm deg} + D_3 + D_{\rm cor}$
also constitute a basis-invariant contribution. Importantly, this
contribution includes terms arising purely from fluctuations. In
contrast to $D_1$ and $D_2$, they involve all states throughout the
Brillouin zone. For instance, $D_3$ is obtained from~\cite{Julku2021b}
\begin{equation}
  \begin{aligned}
    &D_{3,\mu\nu} = \frac{1}{2N}\sum_{\vec{k}\neq 0}\sum_{mm'ss'}
    ss'\frac{n_B[sE_m(s\vec{k})] -
      n_B[s'E_{m'}(s'\vec{k})]}{s'E_{m'}^+(s'\vec{k}) -
      sE_{m}(s\vec{k})}\times\\
    \times&\big[
      \bra{m'^{s'}(\vec{k})}\partial_{\mu}H_B(\vec{k})\ket{m^s(\vec{k})}
      \bra{m^s(\vec{k})}\partial_{\nu}H_B(\vec{k})\ket{m'^{s'}(\vec{k})}
      \\
      -&
      \bra{m'^{s'}(\vec{k})}\gamma_z\partial_{\mu}H_B(\vec{k})\ket{m^s(\vec{k})}
      \bra{m^s(\vec{k})} \gamma_z\partial_{\nu}H_B(\vec{k})
      \ket{m'^{s'}(\vec{k})}. 
  \end{aligned}
\end{equation}
Mathematically, $D_3$ is very similar to the geometric part of the
superfluid weight in a fermionic superconductor where Cooper pairs
have a nonzero center-of-mass momentum $\vec{k_c}$ (i.e. an FFLO
state~\cite{Larkin1964,Fulde1964}). In a fermionic BCS superconductor
($\vec{k_c}=0$) on an isolated flat band, the superfluid weight is
proportional to the minimal integrated quantum
metric~\cite{Peotta2015,Huhtinen2022}. However, no similar
relationship is known in FFLO states with arbitrary
$\vec{k_c}$. Similarly, we cannot obtain an explicit expression for
$D_3$ for a general $\vec{k_c}$. We note that by itself, $D_3$ is not
basis-independent. The addition of the other contributions in $D_3'$
restores the basis invariance of the superfluid weight.

In the kagome lattice (Fig.~\ref{fig.kagome}), we find that $D_3'$ is
positive only at very small interactions, and turns negative around $U\approx 0.13t$. In general, the sign of $D_3'$ will depend on the model parameters. This is again similar to the geometric contribution to the
superfluid weight in FFLO phases (or pair density waves) in fermionic
systems, which is known to change sign under certain
conditions~\cite{Kitamura2022b,Jiang2023,Chen2023}. For the following, the
important points to keep in mind are that $D_1+D_2^{\rm ndeg}$, which
is typically the largest positive contribution to the superfluid
weight, depends on the geometry of Bloch functions at $\vec{k_c}$
only, while $D_3'$, which can be positive or negative, involves Bloch
states throughout the Brillouin zone. The disproportionate role played
by the quantum geometry at $\vec{k_c}$, not only in the superfluid
weight but also other quantities such as the speed of sound~\cite{Julku2021a},
has important consequences for the stability of flat-band superfluidity,
as we will see in the following.  

\section{Stability or instability of flat-band superfluids}

For a stable flat-band superfluid, we need at least a non-vanishing
condensate fraction, a non-vanishing speed of sound in all directions,
and a positive definite superfluid weight. For all of these
conditions, the condensate quantum metric $\mathcal{M}^D(\vec{k_c})$ plays an
important role.

As shown in Refs.~\onlinecite{Julku2021a,Julku2023}, the excitation
fraction in the zero-interaction limit is given by 
\begin{equation}
  \lim_{U\to 0}n_{\rm ex} = \frac{1}{N}\sum_{\vec{k}}\frac{1-d(\vec{k})}{2d(\vec{k})},
\end{equation}
where $d(\vec{k})=\sqrt{1-|\alpha(\vec{k})|^2}$, $\alpha(\vec{k}) =
M\sum_{\alpha}
\braket{\overline{n}_{\vec{k_c}+\vec{k}}}{\alpha}\braket{\alpha}{\varphi_0}\braket{\varphi_0^*}{\alpha}\braket{\alpha}{\overline{n}^*_{\vec{k_c}-\vec{k}}}$
is the condensate quantum distance. The condensate quantum metric
gives the lowest order contribution to the condensate quantum
distance,
$d^2(\vec{k})=2\sum_{\mu\nu}k_{\mu}k_{\nu}\mathcal{M}^D_{\mu\nu}(\vec{k_c})+\mathcal{O}(\vec{k}^3)$
(see Appendix~\ref{app.distance}). A vanishing condensate quantum
metric will thus often lead to a very large $n_{\rm ex}$, and
destabilize the condensate. While we focus on the properties of
$\mathcal{M}^D(\vec{k_c})$ and how it relates to physical quantities, it is
important to stress that $d(\vec{k})$ needs to be nonzero at all
$\vec{k}\neq 0$, which might not be the case even if
$\mathcal{M}^D(\vec{k_c})$ is nonzero. The conditions formulated below
in terms of the condensate quantum metric are therefore merely
necessary for stable condensation and superfluidity.

To lowest order in the interaction, the lowest-amplitude eigenvalues
of $\gamma_zH_B(\vec{k})$ are
$\pm E_{\overline{n}}(\vec{k})=\pm(Un_0/M)d(\vec{k})$. The speed of sound
is given by the slope of this mode at $\vec{k_c}$, and is thus
determined by the condensate quantum metric. For the speed of sound
to be non-vanishing in all directions, we need $\mathcal{M}^D(\vec{k_c})$ to be
positive definite, which requires ${\rm det}[\mathcal{M}^D(\vec{k_c})]>0$. We emphasize that considering the determinant is necessary in systems with more than one spatial dimension, since we need to ensure that there are no vanishing eigenvalues, which would indicate a vanishing speed of sound in some direction.

In the case of a single-band condensate with uniform densities, the
condensate quantum metric is also directly proportional to the Hessian
matrix of $E_{\rm MF}(\vec{k}) = n_0(\epsilon_0-\mu) + (Un_0^2/2)
\sum_{\alpha}|\braket{\alpha}{\overline{n}_{\vec{k}}}|^4$ at $\vec{k_c}$:
\begin{equation}
  \begin{aligned}
    \frac{\partial^2 E_{\rm MF}(\vec{k})}{\partial k_i\partial k_j}\bigg|_{\vec{k}=\vec{k_c}} =&
    Un_0\sum_{\alpha}\partial_{i}(|\braket{\alpha}{\varphi_0}|^2)\partial_j(|\braket{\alpha}{\varphi_0}|^2)
    \\
    =& \frac{2Un_0}{M}\mathcal{M}^D_{ij}(\vec{k_c}).
  \end{aligned}
\end{equation}
where we have used that
$\sum_{\alpha}|\braket{\alpha}{\varphi_0}|^2\partial_i\partial_j(|\braket{\alpha}{\varphi_0}|^2)=0$
when $|\braket{\alpha}{\varphi_0}|^2=1/M$ and that
$\partial_i(|\braket{\alpha}{\varphi_0}|^2) = 2{\rm
  Re}[\braket{\partial_i\varphi_0}{\alpha}\braket{\alpha}{\varphi_0}]$. The
momentum $\vec{k_c}$ is picked as a minimum of $E_{\rm MF}(\vec{k})$; the
stability of this minimum is determined by the condensate quantum
metric.

Before even considering the superfluid weight, $\mathcal{M}^D(\vec{k_c})$
therefore plays an essential role for the stability of a flat-band
superfluid. As shown above, it determines only part of the superfluid
weight, $D_1+D_2^{\rm ndeg}$. However, because $\mathcal{M}^D(\vec{k_c})$ needs
to be positive definite for a stable superfluid, $D_1+D_2^{\rm ndeg}$
also needs to be positive definite: under our assumptions, a superfluid cannot be stabilized by the other contributions $D_3'$ if
$D_1+D_2^{\rm ndeg}$ is not positive definite, even though $D_3'$ can
in principle be positive. We will now examine when the necessary
condition ${\rm det}[\mathcal{M}^D(\vec{k_c})]>0$ can or cannot be met.

Let us first focus on models where $H_{\vec{k_c}}$ can be made
real. This is of course the case for all models with $C_2\mathcal{T}$,
where there exists a basis where $H_{\vec{k}}$ is real at all
momenta. It is also the case for all two-band models~: any complex
phase in the off-diagonal components of $H_{\vec{k_c}}$ can be
canceled by a basis transformation, which corresponds to a unitary
transformation $H_{\vec{k}}\to
V_{\vec{k}}^{\vphantom{\dag}}H_{\vec{k}}V_{\vec{k}}^{\dag}$, where $V_{\vec{k}}={\rm
  diag}(1,e^{i\vec{k}\cdot\delta})$ and
$\delta$ is an arbitrary orbital position shift. In a similar vein,
the complex phases of off-diagonal components of $H_{\vec{k_c}}$ can
always be canceled in bipartite lattices where the smaller sublattice
contains only one orbital, e.g. the Lieb lattice. Additionally, in systems with time-reversal
symmetry $H_{\vec{k}}=H_{-\vec{k}}^*$, $H_{\vec{k_c}+\vec{k}}^* =
H_{\vec{k_c}-\vec{k}}$ for time-reversal invariant momenta
$\vec{k_c}$, i.e. $H_{\vec{k_c}}$ can be chosen real.  

In a basis where $H_{\vec{k_c}}$ is real, all Bloch functions at
$\vec{k_c}$ can be picked real, and using Eq.~\eqref{eq.md_reformulated}, we can
rewrite $\mathcal{M}^D(\vec{k_c})=2\sum_{\alpha}\sum_{m>\overline{n}} {\rm
  Re}[\braket{\partial_{\mu}\varphi_0}{m_{\vec{k_c}}}] {\rm
  Re}[\braket{m_{\vec{k_c}}}{\partial_{\nu}\varphi_0}]$ (note that we
assume only $H_{\vec{k_c}}$ is real, and
$\ket{\partial_{\mu}\varphi_0}$ might still be complex). In a
two-dimensional system, the
determinant of $\mathcal{M}^D(\vec{k_c})$ is then
\begin{equation}
  \begin{aligned}
    {\rm det}[\mathcal{M}^D(\vec{k_c})] =
    4\sum_{m,l>\overline{n}}\big[&
    {\rm Re}[\braket{\partial_x\varphi_0}{m_{\vec{k_c}}}]
    {\rm Re}[\braket{m_{\vec{k_c}}}{\partial_x\varphi_0}] \times\\
    \times&
      {\rm Re}[\braket{\partial_y\varphi_0}{l_{\vec{k_c}}}]
      {\rm Re}[\braket{l_{\vec{k_c}}}{\partial_y\varphi_0}]\\
      -&
      {\rm Re}[\braket{\partial_x\varphi_0}{m_{\vec{k_c}}}]
    {\rm Re}[\braket{l_{\vec{k_c}}}{\partial_x\varphi_0}]\times\\
    \times&
      {\rm Re}[\braket{\partial_y\varphi_0}{l_{\vec{k_c}}}]
      {\rm Re}[\braket{m_{\vec{k_c}}}{\partial_y\varphi_0}]
    \big]
  \end{aligned}
\end{equation}
It is immediately evident that for a two-band model, where there is
only one band above the lowest band, ${\rm
  det}[\mathcal{M}^D(\vec{k_c})]=0$ (recall that
$H_{\vec{k_c}}$ can always be made real in a two-band model). This means that in
two-band flat-band models where there exists a uniformly distributed
Bloch function, superfluidity is impossible in two dimensions within
standard Bogoliubov theory. Based on a similar reasoning, we can
determine that superfluidity in a two-band model is also impossible in
three dimensions. In fact, superfluidity in three dimensions is not
even possible in a three-band model if $H_{\vec{k_c}}$ can be made
real. This means that, for instance, in flat-band systems with
$C_2\mathcal{T}$ symmetry, the total number of bands needs to exceed
the number of spatial dimensions for flat-band superfluidity to be possible. 

When $\vec{k_c}$ is a time-reversal invariant momentum (TRIM), e.g. the $\Gamma$
point, $M$ point, etc., the Hamiltonian can be made real at
$\vec{k_c}$ whenever the system has time-reversal symmetry, as
mentioned previously. Additionally, when
$H_{\vec{k_c}-\vec{k}}=H_{\vec{k_c}+\vec{k}}^*$, the derivatives of
Bloch states at $\vec{k_c}$ are purely imaginary,
$\ket{\partial_{\mu}n_{\vec{k_c}}} =
-\ket{\partial_{\mu}n_{\vec{k_c}}^*}$. Since the Hamiltonian at
$\vec{k_c}$ is real, this implies that ${\rm
  Re}[\braket{m_{\vec{k_c}}}{\partial_{\mu}\varphi_0}]=0$. There can
therefore be no superfluidity in a single flat band at a TRIM where the
Bloch state has uniform densities. Note that this result assumes that
there are no band touchings at $\vec{k_c}$: a flat-band superfluid
could in principle be stabilized at a TRIM in the presence of
time-reversal symmetry by a band touching. For instance, in the kagome lattice, condensation at the $K$-point is most favorable, but a metastable superfluid at the $\Gamma$ point is in principle possible within Bogoliubov theory~\cite{You2012}.

When there exists no basis in which $H_{\vec{k_c}}$ is real,
$\mathcal{M}^D_{\mu\mu}(\vec{k_c})$ is bounded from above by the quantum metric,
as seen from Eq.~\eqref{eq.inequality}. Superfluidity can therefore
only occur for condensates at points where the diagonal components of the quantum
metric are nonzero. In systems with $C_2\mathcal{T}$, $\mathcal{M}^D(\vec{k_c})$
is equal to the quantum metric in a particular basis, and superfluidity
cannot occur where ${\rm det}[\mathcal{M}^0(\vec{k_c})]=0$, even if
all diagonal components are nonzero. Because we consider the quantum
metric at only one momentum, there can be models with a highly
geometrically nontrivial flat band, in the sense of a large integrated
quantum metric, that cannot host a stable superfluid. This includes,
for instance, the two-band checkerboard models in Ref.~\onlinecite{Rhim2019}
featuring singular band touchings~\cite{Rhim2021}: even though the
integrated quantum metric diverges due to the band touching,
condensation and superfluidity remain impossible.

So far, we have not considered the impact of $D_3'$. As mentioned
previously, a positive $D_3'$ cannot salvage superfluidity if
$D_1+D_2^{\rm ndeg}$ vanishes. However, since $D_3'$ can also be
negative, it can be that ${\rm det}[D_1+D_2^{\rm ndeg}]\propto{\rm
  det}[\mathcal{M}^D(\vec{k_c})]>0$ is not sufficient for a positive
definite superfluid weight. In two dimensions, when $D_3'$ is negative
semidefinite, a positive definite $D$ requires the following
conditions:
\begin{align}
    &{\rm Tr}[D_1+D_2^{\rm ndeg}] > -{\rm Tr}[D_3'], \\
    &{\rm det}[D_1+D_2^{\rm ndeg}] > {\rm det}[D_3'](1+{\rm
      Tr}[(D_1+D_2^{\rm ndeg})D_3'^{-1}]). 
\end{align}
When $D_3'=-\overline{D_3}'\mathbb{1}$, we need ${\rm
  det}[D_1+D_2] > \overline{D_3}'^2$. This means that ${\rm
  Tr}[\mathcal{M}^D(\vec{k_c})]$ and ${\rm det}[\mathcal{M}^D(\vec{k_c})]$
do not only need to be non-zero, but also large enough to overcome the
effect of fluctuations. Since $D_3'$ depends on the geometry of states
throughout the Brillouin zone, nontrivial geometry at momenta other
than $\vec{k_c}$ can be detrimental to superfluidity, in
contrast to fermionic systems where nontrivial quantum geometry at any
momentum is usually beneficial to superconductivity.

We have focused on condensation in a Bloch state with uniform
densities on a non-degenerate flat band. As explained above, several
conditions can destabilize the superfluid under these
assumptions. Breaking any of these assumptions could, in principle, make
superfluidity easier to achieve. For instance, we have not considered
condensation in degenerate flat bands, which could be particularly
interesting due to the possibility of fragile
topology~\cite{Po2018,Bouhon2019,Bradlyn2019}. Again, in such systems,
$C_2\mathcal{T}$ symmetry guarantees that all Bloch functions at all
momenta can be made real in some basis. In contrast to the single-band
case, however, $\ket{\varphi_0}$ can be complex, since it is a linear
combination of several Bloch states. In Ref.~\onlinecite{realspace},
it is shown that condensation in a Bloch state $\ket{\varphi_0}$ is
impossible if a Bloch state $R\ket{\varphi_0}$, with $R$ a real
diagonal matrix, exists at the same energy and momentum. It follows
immediately that condensation in a degenerate flat band with
$C_2\mathcal{T}$ symmetry is only possible if $\ket{\varphi_0}$ itself
is complex: otherwise, we could obtain a flat-band Bloch state
$R\ket{\varphi_0}$ by simply adding a Bloch function of the degenerate
flat band with $\braket{n_{\vec{k_c}}}{\varphi_0}<1$ to
$\ket{\varphi_0}$. However, a complex $\ket{\varphi_0}$ is not a
sufficient condition for condensation, and in the presence of
degeneracy, the minima of $E_{\rm MF}$ are often unstable; this is the
case for the kagome-III
model~\cite{Balents2002,Bergman2008,Rhim2019,Peri2021}, which features
a lowest degenerate flat band characterized by fragile topology. In
this model, $\ket{\varphi_0}$ needs to be complex in the basis where
all Bloch functions are real, but all minima of $E_{\rm MF}$ are
unstable, rendering condensation impossible (see Appendix~\ref{app.kagomeiii}). In contrast to fermionic
superconductors, where topological invariants provide lower bounds for
the superfluid
weight~\cite{Peotta2015,Tovmasyan2016,Xie2020,Herzog-Arbeitman2022b,Prijon2025,Yu2025b},
nontrivial topology does not guarantee that condensation and superfluidity is possible
in a bosonic system.

\section{Discussion}

By studying the dependence of the superfluid weight on quantum
geometric quantities, we have shown that the geometry of Bloch states
at the condensation momentum, and especially the condensate quantum
metric, plays an important role in stabilizing a flat-band
superfluid. Because the quantum geometry at one momentum disproportionately influences
physical properties such as the superfluid weight and the speed of
sound, a stable flat-band superfluid in a bosonic system is much
more difficult to achieve than flat-band superconductivity. For
instance, we uncover that superfluidity of a condensate in a single flat band is unlikely at
a time-reversal invariant momentum, and requires at least $d+1$ total
bands, where $d$ is the dimension of the system, in the presence of
$C_2\mathcal{T}$ symmetry.

The superfluid weight has additional contributions due to fluctuations
which do not relate to the condensate quantum metric at
$\vec{k_c}$. These contributions involve Bloch states at all momenta
and can be negative, meaning they can contribute to destabilizing the
superfluid. In contrast to fermionic superconductors, where nontrivial
quantum geometry is always beneficial, it can actually become
detrimental in bosonic systems if it occurs away from the condensation
momentum.

Considering the relative difficulty of stabilizing flat-band condensation and superfluidity, many simple flat-band models cannot host a stable
Bose-Einstein condensate. The conditions we formulate can help target
models with more potential for stable condensation and superfluidity. Such flat-band
models can also be constructed by starting from a real-space
perspective~\cite{realspace}. These models can be promising not only for the observation of geometry-enabled condensation in systems with
massively degenerate states, but could also host exotic phases such as
trion condensates~\cite{You2012}.

\begin{acknowledgments}
K.-E.H.~gratefully acknowledges support by an ETH Zürich Postdoctoral Fellowship.
M.D. is grateful for support by a PhD Scholarship of the Studienstiftung des deutschen Volkes and the Würzburg-Dresden Cluster of Excellence ctd.qmat (EXC 2147, project-id 390858490).
\end{acknowledgments}

\appendix

\section{Degenerate perturbation theory for $\gamma_zH_B$ \label{app.perturbation}}

At zero interaction, $\gamma_zH_B(\vec{k})$ in band basis is 
\begin{equation}
H_0(\vec{k}) = \begin{pmatrix}
\epsilon_{\vec{k_c}+\vec{k}}-\epsilon_0 & 0 \\
0 & -(\epsilon_{\vec{k_c}-\vec{k}} - \epsilon_0)
\end{pmatrix},
\end{equation}
where $[\epsilon_{\vec{k}}]_{nm} = \epsilon_{\vec{k}}^n\delta_{nm}$
contains single-particle dispersions and $\epsilon_0$ is the energy of the lowest flat band. The matrix $H_0(\vec{k})$ clearly has
eigenvalues $E_n^+(\vec{k}) = \epsilon^n_{\vec{k_c}+\vec{k}}-\epsilon_0$,
$E_n^-(\vec{k}) = \epsilon_0-\epsilon^n_{\vec{k_c}-\vec{k}}$, with
corresponding eigenvectors
$\ket{u_n^+(\vec{k})}\equiv\ket{+}\otimes\ket{n_{\vec{k_c}+\vec{k}}}$
and
$\ket{u_n^-(\vec{k})}\equiv\ket{-}\otimes\ket{n_{\vec{k_c}-\vec{k}}^*}$. These
eigenvectors form an orthonormal basis. When condensation occurs in a
single flat band (labelled $\overline{n}$), $H_0(\vec{k})$ has two
trivial eigenvalues, and there is freedom in picking the basis of the
corresponding eigenspace. We therefore define
\begin{equation}
\begin{aligned}
  |n^+(\vec{k}) \rangle &= |u_n^+(\vec{k}) \rangle, \:\:\:\:
  |n^-(\vec{k}) \rangle 
  = |u_n^-(\vec{k}) \rangle & &\text{for
  }n\neq\overline{n}, \\
  |n^s(\vec{k}) \rangle &=
  a_{+}^{s}(\vec{k})|u_{\overline{n}}^+(\vec{k}) \rangle + 
  a_{-}^{s}(\vec{k})|u_{\overline{n}}^-(\vec{k}) \rangle &
  &\text{for } n=\overline{n},
\end{aligned}
\end{equation}
where $s=\pm$. To obtain bosonic quasiparticles, we require
normalization such that
$\bra{n^s(\vec{k})}\gamma_z\ket{m^{s'}(\vec{k})} =
s\delta_{mn}\delta_{ss'}$. This clearly holds for any
$\ket{n^s(\vec{k})}=\ket{u_n^s(\vec{k})}$ with $n\neq\overline{n}$. On
the other hand, for $\ket{\overline{n}^s(\vec{k})}$, the
coefficients $a_{s}^{s'}(\vec{k})$ need to be picked properly, and the
basis $\{\ket{n^s(\vec{k})}\}$ is generally not an orthonormal basis,
contrary to $\{\ket{u_n^s(\vec{k})}\}$. The condition
$\bra{n^s(\vec{k})}\gamma_z\ket{n^s(\vec{k})} = s$ is impossible to
fulfill if $\bra{n^s(\vec{k})}\gamma_z\ket{n^s(\vec{k})} = 0$ for some
$n,s$, and this indeed occurs at $\vec{k}=0$ for $n=\overline{n}$. At
this particular $\vec{k}$, $\gamma_zH_B(\vec{k})$ is also not
diagonalizable for any $Un_0\neq 0$. For a stable condensate in a
single mode, $\gamma_zH_B(\vec{k})$ should be diagonalizable at all
$\vec{k}\neq 0$.

Under the uniformity assumption $|\braket{\alpha}{\varphi_0}|^2 = 1/M$, we can approximate at low $Un_0$ that $\gamma_z H_B(\vec{k})\approx H_0(\vec{k})+\Delta V(\vec{k})$, with $\Delta = Un_0/M$ and
\begin{equation}
V(\vec{k}) = \begin{pmatrix}
1 & A(\vec{k}) \\
-A^{\dag}(\vec{k}) & -1
\end{pmatrix}.
\end{equation}
Here, $[A(\vec{k})]_{mn} =
\bra{m_{\vec{k_c}+\vec{k}}}\Phi\ket{n_{\vec{k_c}-\vec{k}}^*}$ and
$[\Phi]_{\alpha\beta} = M\braket{\alpha}{\varphi_0}^2$. Within
perturbation theory, we approximate the eigenvalues and eigenstates
of $\gamma_zH_B(\vec{k})\approx H_0(\vec{k})+\Delta V(\vec{k})$ by
$E_n^s(\vec{k})+\Delta E_{n,1}^s(\vec{k})$ and $\ket{n^s(\vec{k})} +
\Delta \ket{n_1^s(\vec{k})}$. In the following, we leave the momentum
dependence implicit.  

\subsection{First order contributions for $n\neq\overline{n}$}

The first order contributions $E_{n,1}^s$ and $\ket{n_1^s}$ fulfill
\begin{equation}
    H_0\ket{n_1^s} + V\ket{n^s} = E_n^s\ket{n_1^s} + E_{n,1}^s \ket{n^s}. 
    \label{eq.sup.first_order}
\end{equation}
For $n\neq\overline{n}$, this implies that
\begin{equation}
    E_{n,1}^s = s\bra{n^s}\gamma_zV\ket{n^s} = s. 
\end{equation}
To first order in $\Delta$, the eigenvalues of $\gamma_zH_B$ corresponding to $n\neq\overline{n}$ are thus $E_n^s+s\Delta$. 

The corresponding states are obtained by multiplying Eq.~\eqref{eq.sup.first_order} on the left by $\bra{u_m^{s'}}\gamma_z$:
\begin{equation}
\begin{aligned}
&\bra{u_m^{s'}}\gamma_zH_0\ket{n_1^s} + \bra{u_m^{s'}}\gamma_zV\ket{n^s} \\
=& E_n^s\bra{u_m^{s'}}\gamma_z\ket{n_1^s} + E_{n,1}^s\bra{u_m^{s'}}\gamma_z\ket{n^s} \\
\Rightarrow & (E_m^{s'}-E_n^s) \bra{u_m^{s'}}\gamma_z\ket{n_1^s} = sE_{n,1}^s\delta_{mn}\delta_{ss'} - \bra{u_m^{s'}}\gamma_zV\ket{n^s}. 
\end{aligned}
\end{equation}
For $n\neq\overline{n}$, this amounts to 
\begin{equation}
\gamma_z\ket{n_1^s} = \sum_{ms'\neq ns} \frac{\bra{u_m^{s'}}\gamma_zV\ket{n^s}}{E_n^s-E_m^{s'}} \ket{u_m^{s'}}. 
\end{equation}
Note that it is important here to use the orthonormal basis $\ket{u_m^{s'}}$, which is the same as $\ket{m^{s'}}$ only when $m\neq\overline{n}$. Plugging in the expressions for $V$ and $\ket{u_m^s}$, we obtain the first order contributions for $n\neq\overline{n}$,
\begin{equation}
\begin{aligned}
    \ket{n_1^+(\vec{k})} &= 
    -\sum_{m} \frac{\bra{m^*_{\vec{k_c}-\vec{k}}}\Phi^{\dag}\ket{n_{\vec{k_c}+\vec{k}}}}{\epsilon_{\vec{k_c}+\vec{k}}^n + \epsilon_{\vec{k_c}-\vec{k}}^m - 2 \epsilon_0} \ket{-}\otimes\ket{m_{\vec{k_c}-\vec{k}}^*}, \\
    \ket{n_1^-(\vec{k})} &= 
    -\sum_{m}
    \frac{\bra{m_{\vec{k_c}+\vec{k}}}\Phi\ket{n^*_{\vec{k_c}-\vec{k}}}}{\epsilon_{\vec{k_c}-\vec{k}}^n
      + \epsilon_{\vec{k_c}+\vec{k}}^m - 2 \epsilon_0}
    \ket{+}\otimes\ket{m_{\vec{k_c}+\vec{k}}}. \label{eq.notin_states}
\end{aligned}
\end{equation}

\subsection{First order contributions for $n=\overline{n}$}

When condensing into a flat band, it is possible to compute both
$E_{\overline{n},1}^s$ and $\ket{\overline{n}_1^s}$
analytically. Since we focus on contributions to the superfluid weight
involving only $\ket{n^s}$ with $n\neq\overline{n}$, our results do
not require knowledge of either. However, because it relates to quantities
such as the speed of sound, we show here how the eigenvalues
$E_{\overline{n}}^s+\Delta E_{\overline{n},1}^s$ are obtained.

Equation~\eqref{eq.sup.first_order} implies that
\begin{align}
&\bra{u_{\overline{n}}^{s'}}\gamma_zV\ket{\overline{n}^s} = E_{\overline{n},1}^s \bra{u_{\overline{n}}^{s'}}\gamma_z\ket{\overline{n}^s} \\
\Rightarrow & \sum_{s''} a_{s''}^{s} \bra{u_{\overline{n}}^{s'}}\gamma_zV\ket{u_{\overline{n}}^{s''}} = E_{\overline{n},1}^s s'a_{s'}^{s}. \label{eq.sup.deg} 
\end{align}

We now define
\begin{equation}
  \widetilde{V}=\begin{pmatrix}
  1 & \alpha(\vec{k}) \\
  -\alpha(\vec{k})^{\dag} & -1
  \end{pmatrix},
\end{equation}
where $\alpha(\vec{k}) =
\bra{\overline{n}_{\vec{k_c}+\vec{k}}}\Phi\ket{\overline{n}_{\vec{k_c}-\vec{k}}^*}$,
and the vector $[\vec{a}^{s}]_{s'} = a^s_{s'}$. Then
Eq.~\eqref{eq.sup.deg} becomes
\begin{equation}
  \gamma_z\widetilde{V}\vec{a}^s =
  E_{\overline{n},1}^s\gamma_z\vec{a}^s. 
\end{equation}
The coefficients $a_{s'}^s$ and the first-order contributions to the
eigenvalues can thus be determined by diagonalizing
$\widetilde{V}$. The two trivial eigenvalues at $U=0$ thus split into
\begin{equation}
  \pm E_n = \pm \frac{Un_0}{M}\sqrt{1-|\alpha(\vec{k})|^2} = \pm
  \frac{Un_0}{M} d(\vec{k}),
\end{equation}
where $d(\vec{k})$ is the condensate quantum distance.

We note that while we focus on condensation in a single band, this
calculation is easily generalized to condensation in a degenerate set
of flat bands $\mathcal{B}$. In that case, the lowest eigenvalues are
obtained by diagonalizing
\begin{equation}
  \widetilde{V}(\vec{k}) = \begin{pmatrix}
    \mathbb{1} & \widetilde{A}(\vec{k}) \\
    -\widetilde{A}^{\dag}(\vec{k}) & -\mathbb{1}, 
  \end{pmatrix}
\end{equation}
where $[\widetilde{A}]_{nm} =
\bra{n_{\vec{k_c}+\vec{k}}}\Phi\ket{m_{\vec{k_c}-\vec{k}}^*}$ with
$n,m\in\mathcal{B}$. This would generally require numerical
diagonalization. In the important special case where all Bloch
functions at $\vec{k_c}$ and $\ket{\varphi_0}$ are real,
$\widetilde{A}(0) = \mathbb{1}$. This signals that at least to first
order in $U$, the degeneracy of the flat-band states is not lifted at
$\vec{k}=0$, and condensation in a single mode is impossible. This is
consistent with the observation made in Ref.~\onlinecite{realspace}
that condensation in a state $\ket{\varphi_0}$ is impossible if
another state $R\ket{\varphi_0}$, with $R$ a real diagonal matrix not
proportional to the identity, exists on the flat band.

\section{Contribution $D_2^{\rm ndeg}$ to the superfluid weight} \label{app.d2}

As stated in the main text, $D_2$ is given by
\begin{equation}
  \begin{aligned}
  &D_{2,\mu\nu} = -n_0\lim_{\vec{k}\to 0} \sum_{ms}\\
  &\frac{\bra{\vec{\varphi_0}}
    \gamma_z\partial_{\mu}H_B(-\vec{k}/2)\ket{m^s(-\vec{k})}
    \bra{m^s(-\vec{k})}
    \gamma_z\partial_{\nu}H_B(-\vec{k}/2)\ket{\vec{\varphi_0}}}{E_m^+(-s\vec{k})}.
  \end{aligned}
\end{equation}

We split this expression into two parts, with $D_2^{\rm ndeg}$
containing those terms where $m\neq\overline{n}$, and $D_2^{\rm deg}$
those with $m=\overline{n}$. This work focuses on the former contribution.

\subsection{Low interactions}

At low interactions, we determine $D_2^{\rm ndeg}$ by simply plugging
in $\ket{n^s}=\ket{u_n^s}+\Delta \ket{n_1^s}$, with $\ket{n_1^s}$
given by Eq.~\eqref{eq.notin_states}, in the expression for $D_2^{\rm
  ndeg}$. Because we do not consider the states $m=\overline{n}$, the
limit $\vec{k}\to 0$ can be taken without issue, and to first order in $U$,
\begin{equation}
  \begin{aligned}
    &\lim_{\vec{k}\to
      0}\bra{\vec{\varphi_0}}\gamma_z\partial_{\mu}H_B(-\vec{k}/2)\ket{m^+(-\vec{k})}
    \\
    =& \bra{\varphi_0}\partial_{\mu}H_{\vec{k_c}}\ket{m_{\vec{k_c}}}
    \\
    -& \frac{Un_0}{M} \sum_{l} \frac{\bra{l_{\vec{k_c}}^*} \Phi^{\dag}
    \ket{m_{\vec{k_c}}}}{ \epsilon_{\vec{k_c}}^l +
      \epsilon_{\vec{k_c}}^m - 2\epsilon_0}
    \bra{\varphi_0^*}\partial_{\mu}H_{\vec{k_c}}^*\ket{l_{\vec{k_c}}^*},
    \\
    &\lim_{\vec{k}\to
      0}\bra{\vec{\varphi_0}}\gamma_z\partial_{\mu}H_B(-\vec{k}/2)\ket{m^-(-\vec{k})}
    \\
    =& \bra{\varphi_0^*}\partial_{\mu}H_{\vec{k_c}}^*\ket{m_{\vec{k_c}}^*}
    \\
    -& \frac{Un_0}{M} \sum_{l} \frac{\bra{l_{\vec{k_c}}} \Phi
    \ket{m_{\vec{k_c}}^*}}{ \epsilon_{\vec{k_c}}^l +
      \epsilon_{\vec{k_c}}^m - 2\epsilon_0}
    \bra{\varphi_0}\partial_{\mu}H_{\vec{k_c}}\ket{l_{\vec{k_c}}},
  \end{aligned}
\end{equation}
Since to first order in $U$, $E_m^s(\vec{k})\approx
s(\epsilon_{\vec{k_c}+s\vec{k}} -\epsilon_0 + Un_0/M)$, we get
\begin{equation}
  \begin{aligned}
    D_{2,\mu\nu}^{\rm ndeg} &= -n_0\sum_{m\neq\overline{n}}
    \frac{ \bra{\varphi_0}
      \partial_{\mu}H_{\vec{k_c}}\ket{m_{\vec{k_c}}}
      \bra{m_{\vec{k_c}}} \partial_{\nu}H_{\vec{k_c}}\ket{\varphi_0} +
      \mu\leftrightarrow\nu}{\epsilon_{\vec{k_c}}^m-\epsilon_0} \\
    +& \frac{Un_0^2}{M}\sum_{m\neq\overline{n}}\sum_l
    \frac{\bra{l_{\vec{k_c}}} \Phi \ket{m_{\vec{k_c}}^*}
    }{\epsilon_{\vec{k_c}}^l + \epsilon_{\vec{k_c}}^m -
      2\epsilon_0}\times \\
    \times&\frac{ \bra{\varphi_0}\partial_{\mu}H_{\vec{k_c}}
      \ket{l_{\vec{k_c}}} \bra{m_{\vec{k_c}}^*}
      \partial_{\nu}H_{\vec{k_c}}^* \ket{\varphi_0^*} + \mu\leftrightarrow\nu }{\epsilon_{\vec{k_c}}^m-\epsilon_0} \\
    +& \frac{Un_0^2}{M}\sum_{m\neq\overline{n}}\sum_l
    \frac{\bra{l_{\vec{k_c}}^*} \Phi^{\dag} \ket{m_{\vec{k_c}}}
    }{\epsilon_{\vec{k_c}}^l + \epsilon_{\vec{k_c}}^m -
      2\epsilon_0}\times \\
    \times&\frac{ \bra{\varphi_0^*}\partial_{\mu}H_{\vec{k_c}}^*
      \ket{l_{\vec{k_c}}^*} \bra{m_{\vec{k_c}}}
      \partial_{\nu}H_{\vec{k_c}} \ket{\varphi_0} +
      \mu\leftrightarrow\nu }{\epsilon_{\vec{k_c}}^m-\epsilon_0} \\
    +& \frac{Un_0^2}{M} \sum_{m\neq\overline{n}}
    \frac{\bra{\varphi_0}\partial_{\mu}
      H_{\vec{k_c}}\ket{m_{\vec{k_c}}} \bra{m_{\vec{k_c}}}
      \partial_{\nu}H_{\vec{k_c}} \ket{\varphi_0} +
      \mu\leftrightarrow\nu}{ (\epsilon_{\vec{k_c}}^m-\epsilon_0)^2}. 
  \end{aligned}
\end{equation}

Now, using that
$\bra{\varphi_0}\partial_{\mu}H_{\vec{k_c}}\ket{m_{\vec{k_c}}} =
(\epsilon_0-\epsilon_{\vec{k_c}}^m)\braket{\partial_{\mu}\varphi_0}{m_{\vec{k_c}}}$,
we obtain
\begin{equation}
  \begin{aligned}
  D_{2,\mu\nu}^{\rm ndeg} &=
  -n_0\sum_{m\neq\overline{n}}(\epsilon_{\vec{k_c}}^m-\epsilon_0)
  [\braket{\partial_{\mu}\varphi_0}{ m_{\vec{k_c}}}
    \braket{m_{\vec{k_c}}}{\partial_{\nu}\varphi_0} +
    \mu\leftrightarrow\nu] \\
  +& \frac{2Un_0^2}{M}\sum_{m,l\neq\overline{n}}
  \frac{\epsilon_{\vec{k_c}}^l-\epsilon_0}{\epsilon_{\vec{k_c}}^l+\epsilon_{\vec{k_c}}^m-2\epsilon_0}\times \\
       \times&{\rm Re}\left[ \braket{\partial_{\mu}\varphi_0}{l_{\vec{k_c}}}
         \bra{l_{\vec{k_c}}} \Phi\ket{m_{\vec{k_c}}^*}
         \braket{m_{\vec{k_c}}^*}{\partial_{\nu}\varphi_0^*} +
         \mu\leftrightarrow\nu \right]\\
       +& \frac{Un_0^2}{M} \sum_{m\neq\overline{n}}
       \left[\braket{\partial_{\mu}\varphi_0}{m_{\vec{k_c}}}
         \braket{m_{\vec{k_c}}}{\partial_{\nu}\varphi_0}
         +\mu\leftrightarrow\nu \right] \\
       =& -D_1 +\frac{2Un_0^2}{M}\mathcal{M}^D_{\mu\nu}(\vec{k_c}), \label{eq.d2_derivation}
  \end{aligned}
\end{equation}
with
\begin{equation}
  \begin{aligned}
    \mathcal{M}^D_{\mu\nu}(\vec{k_c}) =& 
    \sum_{m\neq\overline{n}} 
        {\rm Re}\left[\braket{\partial_{\mu}\varphi_0}{m_{\vec{k_c}}}
          \braket{m_{\vec{k_c}}}{\partial_{\nu}\varphi_0}
          \right] \\
        +& \sum_{m,l\neq\overline{n}} {\rm Re}\left[
          \braket{\partial_{\mu}\varphi_0}{l_{\vec{k_c}}}
          \bra{l_{\vec{k_c}}}\Phi \ket{m_{\vec{k_c}}^*}
          \braket{m_{\vec{k_c}}^*}{\partial_{\nu}\varphi_0^*} \right].
  \end{aligned}
\end{equation}
To get to the last line of Eq.~\eqref{eq.d2_derivation}, we have rearranged the second sum using that
\begin{equation}
\begin{aligned}
&\frac{\epsilon_{\vec{k_c}}^l-\epsilon_0}
{\epsilon_{\vec{k_c}}^l + \epsilon_{\vec{k_c}}^m - 2\epsilon_0} 
{\rm Re}[\braket{\partial_{\mu}\varphi_0}{l_{\vec{k_c}}} 
\bra{l_{\vec{k_c}}} \Phi \ket{m_{\vec{k_c}}^*} \braket{m_{\vec{k_c}}^*}{\partial_{\nu}\varphi_0^*}] \\
+& \frac{\epsilon_{\vec{k_c}}^m-\epsilon_0}
{\epsilon_{\vec{k_c}}^m + \epsilon_{\vec{k_c}}^l - 2\epsilon_0} 
{\rm Re}[\braket{\partial_{\mu}\varphi_0}{m_{\vec{k_c}}} 
\bra{m_{\vec{k_c}}} \Phi \ket{l_{\vec{k_c}}^*} \braket{l_{\vec{k_c}}^*}{\partial_{\nu}\varphi_0^*}] \\
=& {\rm Re}[\braket{\partial_{\mu}\varphi_0}{l_{\vec{k_c}}} 
\bra{l_{\vec{k_c}}} \Phi \ket{m_{\vec{k_c}}^*} \braket{m_{\vec{k_c}}^*}{\partial_{\nu}\varphi_0^*}].
\end{aligned}
\end{equation}

For a condensate with uniform densities,
$|\braket{\alpha}{\varphi_0}|^2=1/M$, we can rewrite
$\ket{m_{\vec{k_c}}}\bra{m_{\vec{k_c}}} =
M\sum_{l\neq\overline{n}}\sum_{\alpha}\ket{l_{\vec{k_c}}}\braket{l_{\vec{k_c}}}{\alpha}
\braket{\alpha}{\varphi_0}\braket{\varphi_0}{\alpha}\braket{\alpha}{m_{\vec{k_c}}}
\bra{m_{\vec{k_c}}}$ when $m\neq\overline{n}$. Since
$\bra{l_{\vec{k_c}}}\Phi\ket{m_{\vec{k_c}}^*} =M
\sum_{\alpha}\braket{l_{\vec{k_c}}}{\alpha}\braket{\alpha}{\varphi_0}
\braket{\varphi_0^*}{\alpha} \braket{\alpha}{m_{\vec{k_c}}^*}$,
\begin{equation}
  \begin{aligned}
    \mathcal{M}^D_{\mu\nu}(\vec{k_c}) =
    2M\sum_{m,l\neq\overline{n}}\sum_{\alpha} &{\rm Re}[
      \braket{\partial_{\mu}\varphi_0}{l_{\vec{k_c}}}
      \braket{l_{\vec{k_c}}}{\alpha} \braket{\alpha}{\varphi_0} ] \times \\
    \times&{\rm Re}[ \braket{\varphi_0}{\alpha}
      \braket{\alpha}{m_{\vec{k_c}}}\braket{m_{\vec{k_c}}}{\partial_{\nu}\varphi_0}].
  \end{aligned}
\end{equation}
This expression can be simplified further by using that
$\ket{\varphi_0}\bra{\varphi_0} = \mathbb{1}-\sum_{m\neq\overline{n}}
\ket{m_{\vec{k_c}}}\bra{m_{\vec{k_c}}}$, and that
$\braket{\partial_{\mu}\varphi_0}{\varphi_0} =
-\braket{\varphi_0}{\partial_{\mu}\varphi_0}=-(\braket{\partial_{\mu}\varphi_0}{\varphi_0})^*$. With
these identities, we can rewrite
\begin{equation}
  \mathcal{M}^D_{\mu\nu}(\vec{k_c}) = 2\sum_{\alpha}{\rm
    Re}[\braket{\partial_{\mu}\varphi_0}{\alpha}\braket{\alpha}{\varphi_0}]{\rm
    Re}[\braket{\varphi_0}{\alpha}\braket{\alpha}{\partial_{\nu}\varphi_0}]. 
\end{equation}

\subsection{Arbitrary interaction} \label{app.d2_arbitrary}

The contribution $D_2^{\rm ndeg}$ can be related to the components of
the quantum metric at arbitrary interactions when $H_{\vec{k_c}}$ (and
thus all Bloch states at $\vec{k_c}$), and
the derivatives of Bloch states at $\vec{k_c}$ are  
real. Under these conditions, it is in principle possible to perform
the degenerate perturbation theory used above to infinite order, but
we find it more straightforward to consider the parameter
$\vec{k}$ in $H_B(\vec{k})$ as a perturbation parameter. This is
possible when computing $D_2^{\rm ndeg}$ because the eigenvectors
$\ket{n^s}$ corresponding to $n\neq\overline{n}$ remain well-behaved
even in the limit $\vec{k}\to 0$. The eigenvectors
$\ket{\overline{n}^s}$ and corresponding eigenvalues, on the other
hand, do not remain differentiable with reference to $\vec{k}$, but
they are not needed in the computation of $D_2^{\rm ndeg}$.

This method relies on the fact that when $\ket{\varphi_0}$ is real
with $|\braket{\alpha}{\varphi_0}|^2=1/M$,
\begin{equation}
  \begin{aligned}
    H_B(\vec{k}) =& \begin{pmatrix}
      H_{\vec{k_c}+\vec{k}}-\mu_{\rm eff} & \Delta\mathbb{1} \\
      \Delta\mathbb{1} & H_{\vec{k_c}-\vec{k}}^* -\mu_{\rm eff},
    \end{pmatrix}\\
    =& G_{\vec{k}}
    \begin{pmatrix}
      \epsilon_{\vec{k_c}+\vec{k}}-\mu_{\rm eff} &
      \crea{\mathcal{G}}_{\vec{k_c}+\vec{k}}
      \vec{\Delta}\mathcal{G}_{\vec{k_c}-\vec{k}}^*  \\
      \mathcal{G}_{\vec{k_c}-\vec{k}}^T
      \vec{\Delta}\mathcal{G}_{\vec{k_c}+\vec{k}} &
      \epsilon_{\vec{k_c}-\vec{k}}-\mu_{\rm eff}
    \end{pmatrix}
    G_{\vec{k}}^{\dag},\\
    G_{\vec{k}} =& \begin{pmatrix}
      \mathcal{G}_{\vec{k_c}+\vec{k}} & 0 \\
      0 & \mathcal{G}_{\vec{k_c}-\vec{k}}^*
    \end{pmatrix}. 
  \end{aligned}
\end{equation}
where $\Delta = Un_0/M$, $\mu_{\rm eff} = \epsilon_0-\Delta$ and
$H_{\vec{k}}=\mathcal{G}_{\vec{k}}\epsilon_{\vec{k}}\mathcal{G}_{\vec{k}}^{\dag}$. 

Now, if $H_{\vec{k_c}}$ is real, it is diagonalizable by a real
$\mathcal{G}_{\vec{k_c}}$ such that
$\mathcal{G}_{\vec{k_c}}\mathcal{G}_{\vec{k_c}}^T=\mathbb{1}$. Then at
exactly $\vec{k}=0$,
$H_B=\mathcal{G}_0\widetilde{H_B}\mathcal{G}_0^{\dag}$, with
$\widetilde{H_B}$ consisting of only diagonal blocks:
\begin{equation}
  \widetilde{H_B} = \begin{pmatrix}
    \epsilon_{\vec{k_c}}-\mu_{\rm eff} & \Delta\mathbb{1} \\
    \Delta\mathbb{1} & \epsilon_{\vec{k_c}}-\mu_{\rm eff}
  \end{pmatrix}. 
\end{equation}
This allows us to find the nonzero eigenvalues $\pm E_{n,0}$ and
corresponding eigenvectors $\ket{n_0^{\pm}}$ of
$\gamma_zH_B(0)$ regardless of $\Delta$. While $\gamma_zH_B(0)$ is not
actually diagonalizable, we can remove this difficulty by adding a
small shift $\epsilon$ to the chemical potential, making $H_B$
positive definite. When taking the limit $\epsilon\to 0$, we obtain
the following nonzero eigenvalues and corresponding eigenvectors for
$\vec{k}\to 0$:
\begin{equation}
  \begin{aligned}
  E_{n,0} =& \sqrt{(\epsilon_{\vec{k_c}}^n-\mu_{\rm eff})^2-\Delta^2},
  \\
  \ket{n_0^s} =& \frac{\epsilon_{\vec{k_c}}^n-\mu_{\rm eff} +
    sE_{n,0}}{\mathcal{N}_{n,s}} \ket{+}\otimes \ket{n_{\vec{k_c}}} -
  \frac{\Delta}{\mathcal{N}_{n,s}} \ket{-}\otimes \ket{n_{\vec{k_c}}}, \label{eq.q_pert}
  \end{aligned}
\end{equation}
where $\mathcal{N}_{n,s} = \sqrt{s[(\epsilon_{\vec{k_c}}^n-\mu_{\rm eff} +
    sE_{n,0})^2-\Delta^2]}$. In the following, we shall ommit writing
the shift $\epsilon$ explicitly, but all results should be interpreted
as the limit $\epsilon\to 0$.

Knowing the eigenvalues and eigenvectors of $\gamma_zH_B(\vec{k})$ at
$\vec{k}=0$, we can approximate $H_B(\vec{k})= H_B(0) +
\sum_{\mu}k_{\mu}H_{B,\mu}(0) + \mathcal{O}(\vec{k}^2)$,
$E_n^s(\vec{k})= sE_{n,0} +
\sum_{\mu}k_{\mu}E_{n,1,\mu}^s+\mathcal{O}(\vec{k}^2)$ and
$\ket{n^s(\vec{k})}= \ket{n_0^s} +
\sum_{\mu}k_{\mu}\ket{n_{1,\mu}^s} +
\mathcal{O}(\vec{k}^2)$. Since for $n\neq\overline{n}$, $E_{n,0}>0$,
when taking the limit $\vec{k}\to 0$ in
\begin{equation}
  \begin{aligned}
    &D_{2,\mu\nu}^{\rm ndeg} =
    -n_0\lim_{\vec{k}\to 0} \sum_{m\neq\overline{n}}\sum_s \\
    &\frac{\bra{\vec{\varphi_0}} \gamma_z\partial_{\mu}H_{B}(-\vec{k}/2) \ket{m^s(-\vec{k})}  \bra{m^s(-\vec{k})} \gamma_z\partial_{\nu}H_B(-\vec{k}/2) \ket{\vec{\varphi_0}} } {E_m^+(-s\vec{k})},
  \end{aligned}
\end{equation}
we in fact need to know only the zeroth order terms, since all other
orders will vanish. In other terms,
\begin{equation}
  D_{2,\mu\nu}^{\rm ndeg} = -n_0\sum_{s}\sum_{m\neq\overline{n}}
  \frac{ \bra{\vec{\varphi_0}} \gamma_z\partial_{\mu}H_B(0)
    \ket{m_0^s}
    \bra{m_0^s}\gamma_z\partial_{\nu}H_B(0)\ket{\vec{\varphi_0}}}{E_{m,0}}. 
\end{equation}
For the eigenvectors given in Eq.~\eqref{eq.q_pert},
\begin{equation}
  \bra{\vec{\varphi_0}\gamma_z\partial_{\mu}}H_B(0)\ket{m_0^s} =
  \frac{\epsilon_{\vec{k_c}}^m-\mu_{\rm eff} +
  sE_{m,0} - \Delta}{\mathcal{N}_{m,s}}\bra{\varphi_0}
  \partial_{\mu}H_{\vec{k_c}}\ket{m_{\vec{k_c}}},
\end{equation}
and
\begin{equation}
  \begin{aligned}
    D_{2,\mu\nu}^{\rm ndeg} = -n_0\sum_{s}\sum_{m\neq\overline{n}}
    &\frac{\left( \epsilon_{\vec{k_c}}^m-\mu_{\rm eff} +
      sE_{m,0}-\Delta \right)^2}{ s\left[
        (\epsilon_{\vec{k_c}}^m-\mu_{\rm eff} + sE_{m,0})^2-\Delta^2
        \right]E_{m,0}}\times\\
    \times&
    \bra{\varphi_0}\partial_{\mu}H_{\vec{k_c}}\ket{m_{\vec{k_c}}}
    \bra{m_{\vec{k_c}}} \partial_{\nu}H_{\vec{k_c}}\ket{\varphi_0}.
  \end{aligned}
\end{equation}
This can be rewritten in the form
\begin{equation}\begin{aligned}
  D_{2,\mu\nu}^{\rm ndeg} &= -2n_0\sum_{m\neq\overline{n}}
  \frac{(\epsilon_{\vec{k_c}}^m-\epsilon_0)^2}{\epsilon_{\vec{k_c}}^m-\epsilon_0+2\Delta}
  \times\\
  \times& \left[ \braket{\partial_{\mu}\varphi_0}{m_{\vec{k_c}}}
    \braket{m_{\vec{k_c}}}{\partial_{\nu}\varphi_0} \right].\label{eq.arbitrary_u} 
\end{aligned}\end{equation}
Note that this equation holds only in a basis where $H_{\vec{k_c}}$,
$\ket{\varphi_0}$, and their derivatives are all real. A similar
derivation can in principle be performed in any basis provided there
exists one where both $H_{\vec{k_c}}$ and $\ket{\varphi_0}$ are real
(which is sufficient for $\widetilde{H_B}$ to consist of only diagonal
blocks) to obtain a general expression valid in any basis.

Eq.~\eqref{eq.arbitrary_u} does not directly involve the quantum
metric itself in general. However, when
$(\epsilon_{\vec{k_c}}^m-\epsilon_0)\equiv E_{\rm gap}$ is
the same for all $m\neq\overline{n}$, i.e. all bands above the flat
band are degenerate at $\vec{k_c}$,
\begin{equation}
  D_{2,\mu\nu}^{\rm ndeg} = -2n_0\frac{E_{\rm gap}^2}{E_{\rm
      gap}+2\Delta}\sum_{m\neq\overline{n}}\braket{\partial_{\mu}\varphi_0}{m_{\vec{k_c}}}\braket{m_{\vec{k_c}}}{\partial_{\nu}\varphi_0},
\end{equation}
i.e. $D_{2,\mu\nu}^{\rm ndeg}$ becomes proportional to the quantum
metric.

Moreover, if $(\epsilon_{\vec{k_c}}^m-\epsilon_0)\gg \Delta$,
$(\epsilon_{\vec{k_c}}^m-\epsilon_0)^2/(\epsilon_{\vec{k_c}}^m-\epsilon_0+2\Delta)\approx
\epsilon_{\vec{k_c}}^m-\epsilon_0-2\Delta$, and
\begin{equation}
  \begin{aligned}
    D_{2,\mu\nu}^{\rm ndeg} =& -n_0\sum_{m\neq\overline{n}}
    (\epsilon_{\vec{k_c}}^m-\epsilon_0) \left[
      \braket{\partial_{\mu}\varphi_0}{m_{\vec{k_c}}}
      \braket{m_{\vec{k_c}}}{\partial_{\nu}\varphi_0} +
      \mu\leftrightarrow\nu \right] \\
    +& \frac{2Un_0^2}{M}\sum_{m\neq\overline{n}} \left[
      \braket{\partial_{\mu}\varphi_0}{m_{\vec{k_c}}}
      \braket{m_{\vec{k_c}}}{\partial_{\nu}\varphi_0} +
      \mu\leftrightarrow\nu \right] 
  \end{aligned}
\end{equation}
Note that this expression coincides with the one obtained at low $U$
in the previous section. 

\section{Positive semidefiniteness of $\mathcal{M}^D$} \label{app.positive}

Let $\vec{v}$ be an arbitrary two-component complex vector,
$\vec{v}=(a,b)^T$. We now consider $m = \vec{v}^{\dag} \mathcal{M}^D
\vec{v}$, and aim to show that $m$ is always positive. It is easily
seen that
\begin{equation}
  \begin{aligned}
    m =& |a|^2 \mathcal{M}_{xx}^D + |b|^2 \mathcal{M}_{yy}^D + 2{\rm Re}[ab^*]
    \mathcal{M}_{xy}^D \\
    =& 2M\sum_{\alpha} aa^* {\rm
      Re}[\braket{\partial_x\varphi_0}{\alpha}\braket{\alpha}{\varphi_0}]{\rm
      Re}[\braket{\varphi_0}{\alpha}\braket{\alpha}{\partial_x\varphi_0}]
    \\
    & \hphantom{2\sum}+ ab^*{\rm
      Re}[\braket{\partial_x\varphi_0}{\alpha}\braket{\alpha}{\varphi_0}]{\rm
      Re}[\braket{\varphi_0}{\alpha}\braket{\alpha}{\partial_y\varphi_0}] \\
    & \hphantom{2\sum}+ ba^*{\rm
      Re}[\braket{\partial_y\varphi_0}{\alpha}\braket{\alpha}{\varphi_0}]{\rm
      Re}[\braket{\varphi_0}{\alpha}\braket{\alpha}{\partial_x\varphi_0}] \\
    & \hphantom{2\sum}+ bb^*{\rm
      Re}[\braket{\partial_y\varphi_0}{\alpha}\braket{\alpha}{\varphi_0}]{\rm
      Re}[\braket{\varphi_0}{\alpha}\braket{\alpha}{\partial_y\varphi_0}]
    \\
    =& 2M\sum_{\alpha} \left| a{\rm
      Re}[\braket{\partial_x\varphi_0}{\alpha}\braket{\alpha}{\varphi_0}]
    +  b{\rm
      Re}[\braket{\partial_y\varphi_0}{\alpha}\braket{\alpha}{\varphi_0}]
    \right|^2, 
  \end{aligned}
\end{equation}
showing that $\mathcal{M}^D$ is indeed positive semidefinite. 

\section{Relation between $\mathcal{M}^D$ and the condensate quantum distance} \label{app.distance}

The basis-invariant condensate quantum distance~\cite{Julku2023}
$d(\vec{q})=\sqrt{1-|\alpha(\vec{q})|^2}$, with
\begin{equation}
  \alpha(\vec{q}) =
  M\sum_{\alpha}\braket{\overline{n}_{\vec{k_c}+\vec{q}}}{\alpha}
  \braket{\alpha}{\varphi_0}
  \braket{\varphi_0^*}{\alpha}\braket{\alpha}{(\overline{n}_{\vec{k_c}-\vec{q}})^*}, 
\end{equation}
determines the low energy
excitations of the condensate. In particular, $E_{\overline{n}}(\vec{k_c}+\vec{q})
= Un_0d(\vec{q})/M$, where $E_{\overline{n}}$ is the eigenvalue of
$\gamma_zH_B$ with the lowest absolute value.

To second order in $\vec{q}$,
\begin{equation}
  \begin{aligned}
    |\alpha(\vec{q})|^2 = 1 -2\sum_{\mu\nu} q_{\mu}q_{\nu} {\rm
      Re}\big[&\braket{\partial_{\mu}\varphi_0}{\partial_{\nu}\varphi_0}
      \\
      +& M
      \sum_{\alpha}\braket{\partial_{\mu}\varphi_0}{\alpha}\braket{\alpha}{\varphi_0} \braket{\varphi_0^*}{\alpha}\braket{\alpha}{\partial_{\nu}\varphi_0^*}
      \big] \\
    = 1-4M\sum_{\mu\nu} q_{\mu}q_{\nu}\sum_{\alpha}&{\rm
      Re}[\braket{\partial_{\mu}\varphi_0}{\alpha}\braket{\alpha}{\varphi_0}]
    {\rm
      Re}[\braket{\varphi_0}{\alpha}\braket{\alpha}{\partial_{\nu}\varphi_0}].
  \end{aligned}
\end{equation}
It follows that up to second order in $\vec{q}$,
$d^2(\vec{q})=4\sum_{\mu\nu}q_{\mu}q_{\nu}\mathcal{M}^D_{\mu\nu}(\vec{k_c})$. 

Note that since $\mathcal{M}^D(\vec{k_c})$ can equivalently be written as
\begin{equation}
  \mathcal{M}^D_{\mu\nu}(\vec{k_c}) =
  \frac{M}{2}\sum_{\alpha}\partial_{\mu}(|\braket{\alpha}{\varphi_0}|^2)
  \partial_{\nu}(|\braket{\alpha}{\varphi_0}|^2), 
\end{equation}
the condensate quantum metric is evidently basis invariant: a basis
change only affects the phase of $\braket{\alpha}{\varphi_0}$, but not
its amplitude.

\section{Minima of $E_{\rm MF}$ in the kagome-III lattice} \label{app.kagomeiii}

\begin{figure}
  \includegraphics[width=\columnwidth]{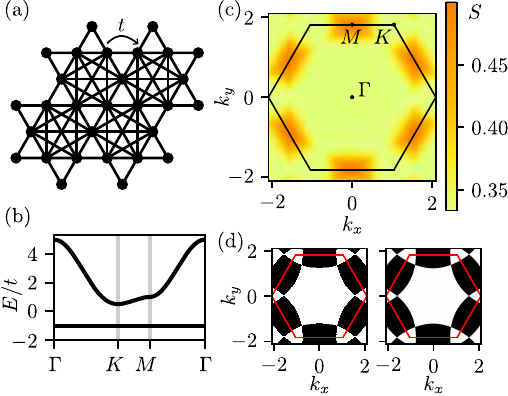}
  \caption{(a) Kagome-III lattice and (b) corresponding
    single-particle band structure. The lowest band at $\epsilon_0=-t$ is
    degenerate. (c) Smallest possible value of
    $S=\sum_{\alpha}|\braket{\alpha}{n_{\vec{k}}}|^4$ obtained from
    numerical minimization at each $\vec{k}$ when
    $\ket{n_{\vec{k}}}$ is a flat-band Bloch function. When considering
    condensation in a flat-band eigenstate, the minima of this sum
    correspond exactly to the minima of $E_{\rm MF} = n_0\epsilon_0-\mu n_0 +
    (Un_0^2/2)S$. The smallest possible value of $S=1/3$ is obtained
    whenever there exists a flat-band Bloch state such that
    $|\braket{\alpha}{n_{\vec{k}}}|^2=1/3$ for all orbitals. (d) Left:
  White indicates all points where the result of the numerical
  minimization fulfills $|S-1/3|<10^{-10}$. Right: White indicates the
  region where
  $|(\cos^2(k_1)+\cos^2(k_2)-\cos^2(k_3))/2\cos(k_1)\cos(k_2)|<1$,
  where constructing a uniform flat-band Bloch state should be
  possible. The results of the numerical minimization are in good
  agreement with this analytical result.}\label{fig.kago-III}
\end{figure}

The single-particle Hamiltonian for the kagome-III lattice, pictured
in Fig.~\ref{fig.kago-III}a, can be written as
\begin{widetext}
\begin{equation}
  H_{\vec{k}} = 2t\begin{pmatrix}
  \cos(2k_3) & \cos(k_1)+\cos(k_2-k_3) & \cos(k_2) + \cos(k_1+k_3) \\
  \cos(k_1)+\cos(k_2-k_3) & \cos(2k_2) & \cos(k_3) + \cos(k_1+k_2) \\
  \cos(k_2)+\cos(k_1+k_3) & \cos(k_3)+ \cos(k_1+k_2) & \cos(2k_1), 
  \end{pmatrix}
\end{equation}
\end{widetext}
where $k_1=k_x$, $k_2 = k_x/2 + \sqrt{3}k_y/2$ and $k_3=k_x/2 -
\sqrt{3}k_y/2$. This lattice features a degenerate lowest flat band,
isolated from a dispersive band (see Fig.~\ref{fig.kago-III}b)

At momentum $\vec{k}$, the eigenspace corresponding to the eigenvalue
$-t$ is spanned by the unnormalized eigenvectors
\begin{equation}
  \begin{aligned}
    \vec{v}_1 =& \left( -\frac{\cos(k_1)}{\cos(k_3)}, 0, 1 \right)^T\\
    \vec{v}_2 =& \left( -\frac{\cos(k_2)}{\cos(k_3)}, 1, 0 \right)^T.
  \end{aligned}
\end{equation}

It is possible to construct a linear combination
$\vec{v}=\vec{v_1}+e^{i\theta}\vec{v_2}$ such that each component has
amplitude $1$ when it is possible to construct a triangle with edge
lengths $1$, $\cos(k_1)/\cos(k_3)$ and $\cos(k_2)/\cos(k_3)$. By the
law of cosines, it is easily determined that this is the case when
\begin{equation}
  \left| \frac{\cos^2(k_1) + \cos^2(k_2) -\cos^2(k_3)}{
    2\cos(k_1)\cos(k_2) } \right| < 1. 
\end{equation}
This condition is true in a large part of the Brillouin zone. The only
regions where it is impossible to construct a uniform-density
flat-band Bloch function are centered around the $M$ points. As a
result, the minima of the mean-field energy are unstable (see
Fig.~\ref{fig.kago-III}c-d).

\end{document}